\documentclass[fleqn,usenatbib]{mnras}
\usepackage{graphicx}	% Including figure files
 \def\deg{^\circ} 
 \def\deg{^\circ}
 \def\/{\over}\def\kms{km s$^{-1}$}     
 \def\co{$^{12}$CO ($J$=1-0) }  \def\coth{$^{13}$CO ($J$=1-0) } \def\coei{C$^{18}$O ($J$=1-0) }   \def\cotw{$^{12}$CO ($J$=1-0) }
 \def\be{\begin{equation}} \def\ee{\end{equation}}
 \def\kms{km s$^{-1}$} \def\Kkms{K \kms }  \def\Ico{I_{\rm CO}} 
  \def\Htwo{H$_2$ }  
\def\({\left(} \def\){\right)} \def\[{\left[} \def\]{\right]}

\def\Tmb{T_{\rm mb}} \def\Tex{T_{\rm ex}} \def\Tx{T_{\rm ex}} 
\def\Tbg{T_{\rm bg}}\def\Tb{T_{\rm B}}
 \def\twco{$^{12}$CO} 
\def\Ico{I_{\rm CO}}
\def\Icotw{I_{\rm ^{12}CO}}
\def\Icoth{I_{\rm ^{13}CO}}

\def\NHtwo{N_{\rm H_2}} \def\Ncoth{N_{\rm ^{13}CO}}
\def\Ycoth{Y_{\rm ^{13}CO}}
\def\uxco{\Htwo [\Kkms]$^{-1}$}
\def\htwo{H$_2$} 
\def\vlsr{v_{\rm lsr}}
\def\Tbcotw{T_{\rm B}({\rm ^{12}CO})} 
\def\Tbcoth{T_{\rm B}({\rm ^{13}CO})} 
\def\Tbcoei{T_{\rm B}({\rm C^{18}O})}
\def\scx{SCD_{12X}} \def\scl{SCD_{\rm 13L} }
\def\scdx{$SCD_{12X}$} \def\scdl{$SCD_{\rm 13L}$ }
\def\scdunit{[${\rm H_2 \ cm^{-2} \ (km \ s^{-1})^{-1}}$]}
\def\xcounit{[\Htwo (K \kms)$^{-1}$]\ }

\def\Xco{X_{\rm ^{12}CO}} 
\def\Xcoth{X_{\rm ^{13}CO,Q}}
\def\Xcross{X_{\rm ^{13}CO, LTE}^{\rm ^{12}CO, Int}}
\def\Xcomod{X^*_{\rm CO}} 
\def\Xgeneral{X_{\rm CO}} 
\def\NthLTE{N_{\rm ^{13}CO, LTE}}
\def\NtwthLTE{N^{^{12}{\rm CO, Int}}_{\rm ^{13}CO, LTE}}
\def\revone{}
\def\revtwo{\textcolor{red}}
\def\revtwo{}
\title[CO-to-H$_2$ Conversion Factor]{CO-to-H$_2$ Conversion and Spectral Column Density in Molecular Clouds: The \revone{Variability} of $X_{\rm CO}$ Factor}
 \author[Y. Sofue and M. Kohno]{
 Yoshiaki Sofue$^{1}$\thanks{E-mail: sofue@ioa.s.u-tokyo.ac.jp}
and Mikito Kohno$^{2, 3}$
\\
$^{1}$Institute of Astronomy, The University of Tokyo, Mitaka, Tokyo 181-0015, Japan\\
$^{2}$Department of Physics, Graduate School of Science, Nagoya University, Furo-cho, Chikusa-ku, Nagoya, Aichi 464-8602, Japan\\
$^{3}$Astronomy Section, Nagoya City Science Museum, 2-17-1 Sakae, Naka-ku, Nagoya, Aichi 460-0008, Japan}
\date{Accepted XXX. Received YYY; in original form ZZZ} 
\pubyear{2020}
\begin{document} 

\maketitle

\begin{abstract}  
\revone{Analyzing the Galactic plane CO survey with the Nobeyama 45-m telescope, we compared the spectral column density (SCD) of \Htwo calculated for \co line using the current conversion factor $\Xco$ to that for \coth line under LTE (local thermal equilibrium) assumption in M16 and W43 regions.
Here, SCD is defined by $dN_{\rm H_2}/dv$ with $\NHtwo$ and $v$ being the column density and radial velocity, respectively.
It is found that the $\Xco$ method significantly under-estimates the \Htwo density in a cloud or region, where SCD exceeds a critical value ($ \sim 3\times 10^{21}$ \scdunit), but over-estimates in lower SCD regions.
We point out that the actual CO-to-\Htwo conversion factor varies with the \Htwo column density or with the CO-line intensity:
It increases in the inner and opaque parts of molecular clouds, whereas it decreases in the low-density envelopes.
However, in so far as the current $\Xco$ is used combined with the integrated $^{12}$CO intensity averaged over an entire cloud, it yields a consistent value with that calculated using the $^{13}$CO intensity by LTE. 
Based on the analysis, we propose a new CO-to-\Htwo conversion relation, 
$\NHtwo^* = \int \Xcomod(\Tb) \Tb\ dv$,
where $\Xcomod (\Tb)=(\Tb/\Tb^*)^\beta \Xco$ is the modified spectral conversion factor as a function of the brightness temperature, $\Tb$, of the \co line, and  $\beta\sim 1-2$ and $\Tb^*=12-16$ K are empirical constants obtained by fitting to the observed data.
The formula corrects for the over/under estimation of the column density at low/high-CO line intensities, and is applicable to molecular clouds with $\Tb \ge 1$ K (\co line rms noise in the data) from envelope to cores at sub-parsec scales (spatial resolution).}
\end{abstract}


\section{Introduction} 

The CO-to-\htwo conversion factor, $\Xco$, has been determined mainly as a statistical average of the ratio of \twco ($J=1-0$)-line luminosity to Virial mass estimated using velocity dispersion and size for a large number of molecular clouds \citep{1987ApJ...319..730S}. 
\revone{Besides the 
(i) Virial method, the factor has been also obtained by various ways, which include those comparing the \cotw line's integrated intensity with 
(ii) dust column density, or optical and infrared extinction ($A_v$ method)  \citep{Lombardi+2008}, 
(iii) thermal dust far infrared emission \citep{Planck+2011,Planck+2015,Okamoto+2017,Hayashi+2019a}, 
(iv) $\gamma$-ray brightness \citep{Bloemen+1986,Abdo+2010,Hayashi+2019}, 
and (v) X-ray shadows \citep{Sofue+2016}.
Thanks to the extensive measurements in the last decades, the $\Xco$ factor appears to be converging to a value of $\Xco \sim 2.0\times 10^{20}$ \uxco in the solar vicinity (see the review by \citealp{2013ARA&A..51..207B}, and the literature therein), suggesting that the factor is a universal constant, which is, however,  dependent on the metal abundance, or on the galacto-centric distance and galaxy types \citep{Arimoto+1996,Leroy+2011}.}

However, because $\Xco$ by the current methods gives an average over a cloud, or the ratio of CO luminosity to independently estimated molecular masses by the various ways, it is not trivial if the same conversion can be applied to local column density inside a cloud, when the cloud is resolved, or a particular region is interested.
\revone{There have been extensive studies in the last decade about the variation of $X_{\rm CO}$ with the column density, CO brightness, and interstellar extinction ($A_v$) (e.g., \citealp{2010A&A...518A..45L, 2007MNRAS.379..674W, Heyer+2009, Lee+2014}). 
It has been shown that the column density derived from the current $\Xco$ method applied to \cotw line intensity overestimates the more reliable column derived from \coth LTE method \citep{Heyer+2009}. 
Furthermore, the fact that the \cotw line is opaque in high-density parts of clouds makes it complicated to evaluate the detailed $\Xco$ in cloud cores, although it is often thought that the LVG (large-velocity gradient) transfer of the CO lines \citep{Scoville+1974} may guarantee the universality of $\Xco$. }

\revone{The most reliable way with minimum conversion process to estimate the \htwo column density would be to employ optically thin lines with similar isotopologue and emission mechanism such as \coth and \coei based on the local thermal equilibrium (LTE) assumption \citep{2008ApJ...679..481P}. }
Thereby, the line intensity is directly related to the column density of the emitting CO molecules, and the conversion from CO column to \htwo column is obtained by simply multiplying the abundance ratio determined independently or given a priori.
\revone{The column density in mass of the cloud can be obtained by multiplying a factor of $\sim 1.4$ \Htwo mass to $\NHtwo$ for the solar abundance ratios in mass among H (0.706), He (0.274) and heavier metals including dust (0.019).} 

We aim at examining how accurately the $X_{\rm CO}$ factor can or cannot estimate the local column density by comparing the values calculated using the optically thick \cotw line combined with the current $\Xco$ and those using the optically thin \coth line under the LTE assumption. 
In this paper, we analyze two Galactic molecular regions which we recently studied in detail in the CO lines: a $0\deg.5 \times 0\deg.5$ region around the Pillars of Creation in M16 centered on G17.0-0.75 at $\vlsr\sim 25$ \kms \citep{2020MNRAS.492.5966S}, and a $3\deg \times 2\deg$ region around the GMC complex associated with the star forming region W43 Main at G30.7-0.08 and $\sim 100$ \kms \citep{Kohno+2020, 2019PASJ...71S...1S}. 
A full and more systematic analysis of the Galactic plane and catalogued GMCs will be a subject for the future. 

The CO data were obtained by the FOREST (i.e., FOur beam REceiver System on the 45-m Telescope: \citealp{2016SPIE.9914E..1ZM}) Unbiased Galactic plane Imaging survey with the Nobeyama 45 m telescope (FUGIN:  \citealp{2017PASJ...69...78U}) project,
\revone{which provided with high-sensitivity, high-spatial and velocity resolution, and wide velocity ($482\ {\rm channels}\times 0.65$ \kms) and field ($40\deg \times 2\deg$ along the Galactic plane from $l=10\deg$ to $50\deg$) coverage by $(l,b,\vlsr:\Tb)$ cubes in the \co, \coth, and \coei lines.
Here, $\Tb$ is the corrected main-beam antenna temperature and is assumed to be equal to the brightness temperature. }

\revone{The full beam width at half maximum of the telescope was $15''$ and $16''$ at the \co and \coth-line frequencies, respectively. 
The effective beam size of the final data cube, convolved with a Bessel $\times$ Gaussian function, was 20\arcsec for $^{12}$CO and 21\arcsec for $^{13}$CO. 
The relative intensity uncertainty is estimated at 10--20\% for $^{12}$CO and 10\% for $^{13}$CO by observation of the standard source M17 SW \citep{2017PASJ...69...78U}. 
The final 3D FITS cube has a voxel size of $(\Delta l, \Delta b, \Delta \vlsr) = (8.5\arcsec, 8.5\arcsec, 0.65 $ $\>$km s$^{-1}$).
The root-mean-square (rms) noise levels are $\sim$1.5 K and $\sim$ 0.9 K for $^{12}$CO and $^{13}$CO in W43, and $\sim 1$ K and $\sim 0.7$ K in M16 region, respectively.}

\section{Basic relations}

The brightness temperature of CO lines, $\Tb$, which is assumed to be equal to the observed main-beam temperature, $\Tmb$, is expressed in terms of the excitation temperature, $\Tex$, and optical depth, $\tau$, by (e.g., \citealp{2008ApJ...679..481P}) 
\begin{equation}
\Tb = T_0
\left(\frac{1}{e^{T_0/\Tex}-1} - \frac{1}{e^{T_0/\Tbg}-1 }\right)
\left(1-e^{-\tau}\right),
\label{Tb}
\end{equation}
where $\Tbg=2.725$ K is the black-body temperature  of the cosmic background radiation, 
$T_0=h \nu/k$ is the Planck temperature with 
$h$ and $k$ being the Planck and Boltzmann constants, rexpectively,
%h=6.62607\times 10^{-27}$ cgs  
%=1.3807\times 10^{-16}$ cgs 
and $\nu$ is the frequency of the line.
Table \ref{tab1} lists the Planck temperature $T_0$ for the three CO lines. 

For the \co line, the molecular gas is assumed to be optically thick, so that the excitation temperature can be measured by observing the brightness temperature of the line through
\begin{equation}
\Tex=T_{0}^{115}\times {\rm ln} \( 1+\frac{T_{0}^{115}}{\Tbcotw_{\rm max}+0.83632}  \)^{-1}\ {\rm K}
\label{Tex}
\end{equation}

{We assume that the molecular gas is in thermal equilibrium and the excitation temperatures of \co, {\coth} and {\coei} lines are equal to each other. Then, the above determined $\Tex$ for {\co} line } can be used to estimate the optical depth of {\coth} and {\coei} lines as
\be
\tau(^{13}{\rm CO})=-{\rm ln} 
\(1-\frac{\Tbcoth_{\rm max}/T_{0}^{110} }{(e^{T_{0}^{110}/\Tex}-1)^{-1}- 0.167667} \)
\ee
 and
\be
\tau({\rm C^{18}O})=-{\rm ln} 
\(1-\frac{\Tbcoei_{\rm max}/T_{0}^{109}}{(e^{T_{0}^{109}/\Tex}-1)^{-1}- 0.169119} \),
\ee
where $T_{0}^{115}$, $T_{0}^{110}$ and $T_{0}^{109}$ represent the Planck temperatures, $T_0=h\nu/k$, at corresponding frequencies, and are listed in table \ref{tab1}.

Here, ${\Tb}_{\rm max}$ is the maximum brightness temperature at the line center in each direction, which is, hereafter, approximated by $\Tb$ at each grid of channel map at a representative velocity fixed for the regions under consideration. 

\begin{table} %%%%%%%%%%%%%%%%%%  
\caption{Planck temperatures of the CO lines, $T_0=h \nu/k$}
\begin{center}
\begin{tabular}{lll} 
\hline 
\hline   
Line & Freq., $\nu$ (GHz)& Planck temp., $T_0$ (K)\\
\hline
 \co  & 115.271204 & $T_{0}^{115}=5.53194$ \\ 
 \coth& 110.20137 & $T_{0}^{110}=5.28864$\\
 \coei& 109.782182 &$T_{0}^{109}=5.26852$\\
\hline    
\end{tabular} 
\label{tab1} 
\end{center}  
\end{table} %%%%%%%%%%%%%%%%%%%

The \Htwo column density using \co line with $\Xco$ factor and that using the \coth line on the LTE assumption are defined through 
\be
\NHtwo(^{12}{\rm CO})=\Xco I_{\rm ^{12}CO},
\label{NHtwoXco} 
\ee
where
\be
I_{\rm ^{12}CO}=\int \Tbcotw dv,
\ee
and
\be
\NHtwo ({\rm ^{13}CO})=\Ycoth \Ncoth,
\label{YN}
\ee
and
$ \Xco =2.0\times 10^{20}$ \xcounit is the widely used conversion factor \citep{2013ARA&A..51..207B}, and 
$\Ycoth=(5.0\pm 2.5)\times 10^5$ 
is the abundance ratio of \Htwo to $^{13}$CO molecules \citep{1978ApJS...37..407D}.
We also adopt $\Ycoth=7.7\times 10^5$  in section \ref{integCD} only for comparison with the result of \cite{Kohno+2020}.  

The column density of $^{13}$CO molecules is given by \citep{2008ApJ...679..481P}
\be
\Ncoth=3.0\times 10^{14} ~ Q ~ I_{\rm ^{13}CO}  ,
\ee
where
\be
Q = Q \left( \Tex,\Tb(^{13}{\rm CO}) \right)
=\frac{\tau}{1-e^{-\tau}}
\frac{1}{ 1-e^{-T_{0}^{110}/\Tex}},
\label{Q}
\ee 
and
\be
I_{\rm ^{13}CO}=\int \Tbcoth dv
\ee
is the integrated intensity of the \coth line.
Thus, Eq. \ref{YN} reduces to
\be
\NthLTE = \Xcoth \ I_{\rm {13}CO} ,
\label{NHtwoLTE}
\ee  
where
\be
\Xcoth
= 1.50 \times 10^{20} Q \ [{\rm H_2 cm^{-2}  (K \ km \ s^{-1})^{-1}]} .
\label{eqXcoth}
\ee
is the conversion factor for the \coth line intensity.

\revone{We further introduce a conversion factor, which relates the supposed "true" \Htwo column density from $^{13}$CO LTE method to the $^{12}$CO line intensity,
\be 
\NtwthLTE=\Xcross \Ico.
\label{eqXnew} 
\ee
This is a cross relation between the $^{12}$CO intensity and $^{13}$CO LTE column.
The "cross" conversion factor, $\Xcross$, can be, in principle, determined by plotting $\NthLTE$ against $\Ico$, but, as discussed later (Fig. \ref{newXco}), it is not practical in the present data. 
}
 
\revone{Including another modified relation discussed later, we have, thus, four methods to estimate the H$_2$ column density by CO line observations.}

\revone{{\bf (1) $\Xco$ method:}  
\Htwo column density is calculated for \co line intensity using Eq. (\ref{NHtwoXco}) with the constant conversion factor $\Xco=2.0\times 10^{20}$ \xcounit.}

\revone{{\bf (2) $\Xcoth$ LTE method:} Excitation temperature, $\Tex$, calculated by Eq. (\ref{Tex}) for $\Tb$ of \co line is used to estimate the \Htwo column by Eq. (\ref{NHtwoLTE}) for \coth line intensity under LTE assumption.}

\revone{{\bf (3) Cross (hybrid) $\Xcross$ method:} $^{12}$CO line intensity is used to obtain a more reliable \Htwo column density using the $\Xcross$ factor through Eq. (\ref{eqXnew}).  }

\revone{{\bf (4) Modified conversion method using $\Xcomod$:} This will be discussed later in section \ref{secmodify}, using the modified conversion factor, $\Xcomod$, as presented by Eqs. (\ref{eqXmod}) and (\ref{eqNmod}).
This is an advantageous method, when we have only \co data as often experienced in large-scale surveys and in extra-galactic CO line observations.
} 

%%%%%%%%%%%%%%%%%%%%%%%%%%%%%%%%%%%%%%%%%%%%%%%%%
\section{Integrated column density}
%%%%%%%%%%%%%%%%%%%%%%%%%%%%%%%%%%%%%%%%%%%%%%%%%%
\label{integCD}

We examine the correlation between column densities calculated by the two methods \revone{in the W43} and M16 Pillar regions using the FUGIN \co and \coth line data.

\subsection{Linear correlation between averaged column densities}
\label{avNHtwo}

\begin{figure}%%%%%%%%%%%%%%%% 
\begin{center}        
\includegraphics[width=8cm]{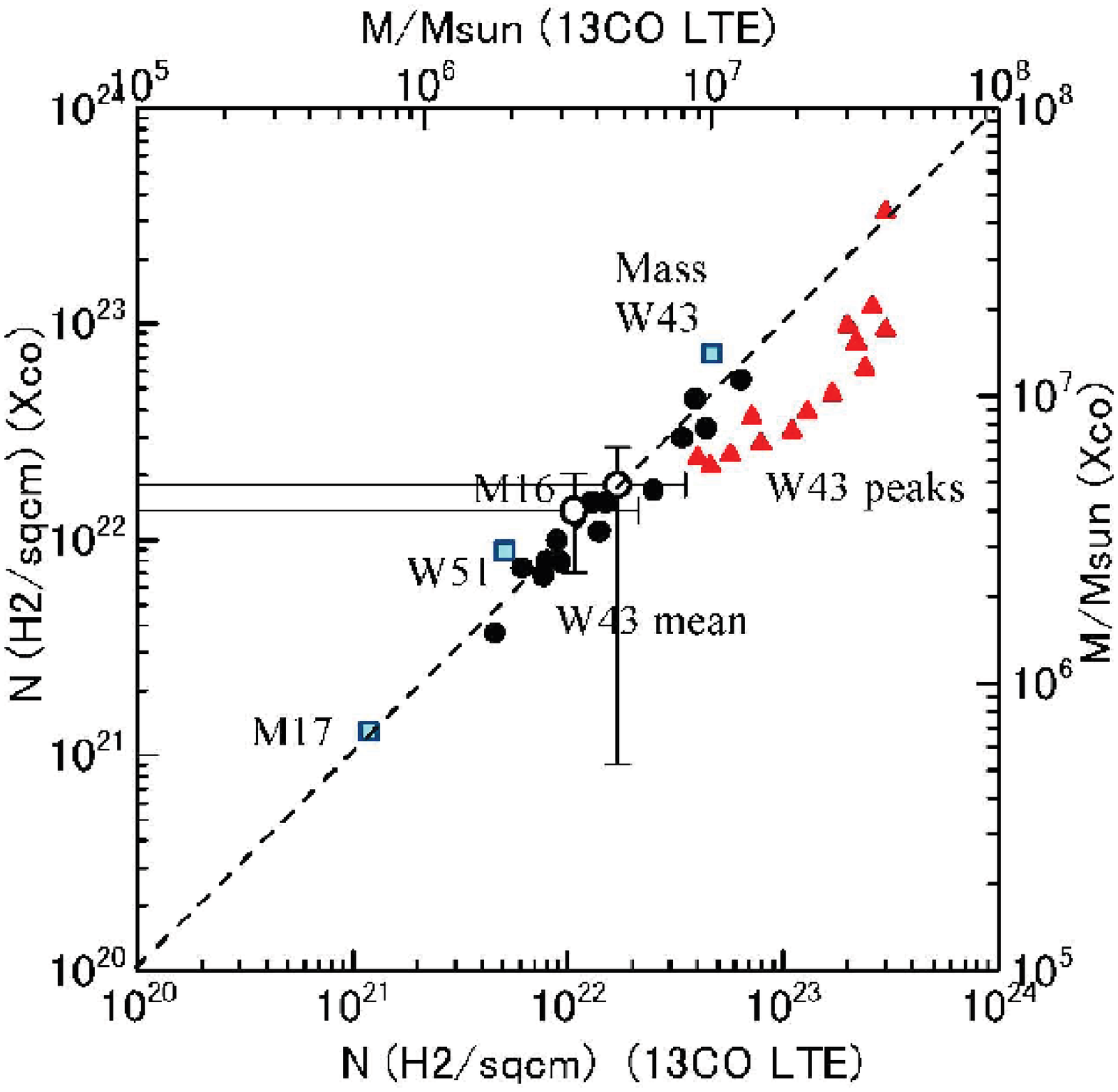}  
\end{center}
\caption{
Column densities of hydrogen molecules calculated by $\Xco$ (\cotw intensity) and LTE (\coth intensity) methods \citep{Kohno+2020}. Filled circles are mean column densities in the identified molecular clouds in the W43 GMC complex, red triangles are the spatial maximum of the integrated area, and rectangles are total H$_2$ masses of three molecular complexes.  The displacement of the peak-intensity values from the linear relation is consistent with that found between \scdx and \scdl. Big circles are average and standard deviation of the plots in figure \ref{NHtwo_m16w43}.} 
\label{NHtwo_all}
\end{figure}

Figure \ref{NHtwo_all} shows plots of the mean \htwo column densities in giant molecular clouds (GMC) in W43 calculated using the $\Xco$ and LTE methods  by \cite{Kohno+2020}. 
Filled circles indicate the mean column densities for individual GMC and cloud components in W43 Main region, and red triangles are those for their intensity peaks. Rectangles show total \htwo masses of W43 and other two molecular complexes. 

The plots show that the mean (averaged) values of the column density in individual molecular clouds calculated using the two different methods are well correlated in a linear fashion. This confirms that $\Xco$ is useful to estimate the total masses of individual molecular clouds. 

On the other hand, the plots of column densities calculated for peak positions of the clouds show significant displacement from the linear relation, as indicated by the red triangles, in the sense that the 
$\Xco$-method under-estimates the local column density at the peak compared to the LTE method. This suggests that the CO-to-H$_2$ conversion from the two method may not be universal in different places in a single cloud.

%%%%%%%%%%%%%%%%%%        
 
\subsection{Non-linear correlation in local column densities}

\begin{figure*}%%%%%%%%%%%%%%%% 
\begin{center}        
\includegraphics[width=14cm]{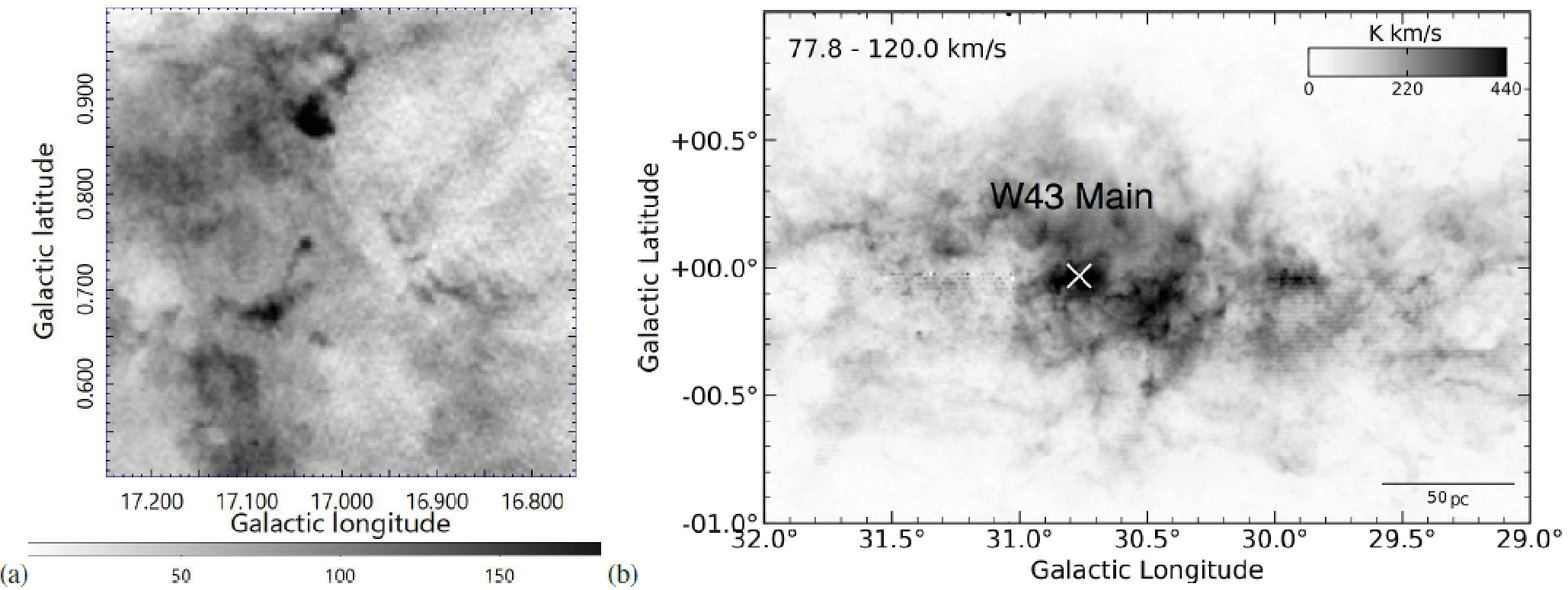}  
\end{center}
\caption{\co integrated intensity maps of (a) M16 (from $\vlsr=20$ to 30 \kms), and (b) W43 (from 77.8 to 120 \kms) regions analyzed in this paper. The X mark indicates W43 Main cloud. Grey scale is in K km s$^{-1}$. The map grids are $8''.5\times 8''.5$.} 
\label{maps}
%\end{figure*}%%%%%%%%%%%%%%%%%%        
%\begin{figure*}%%%%%%%%%%%%%%%% 
\begin{center}         
\includegraphics[width=14cm]{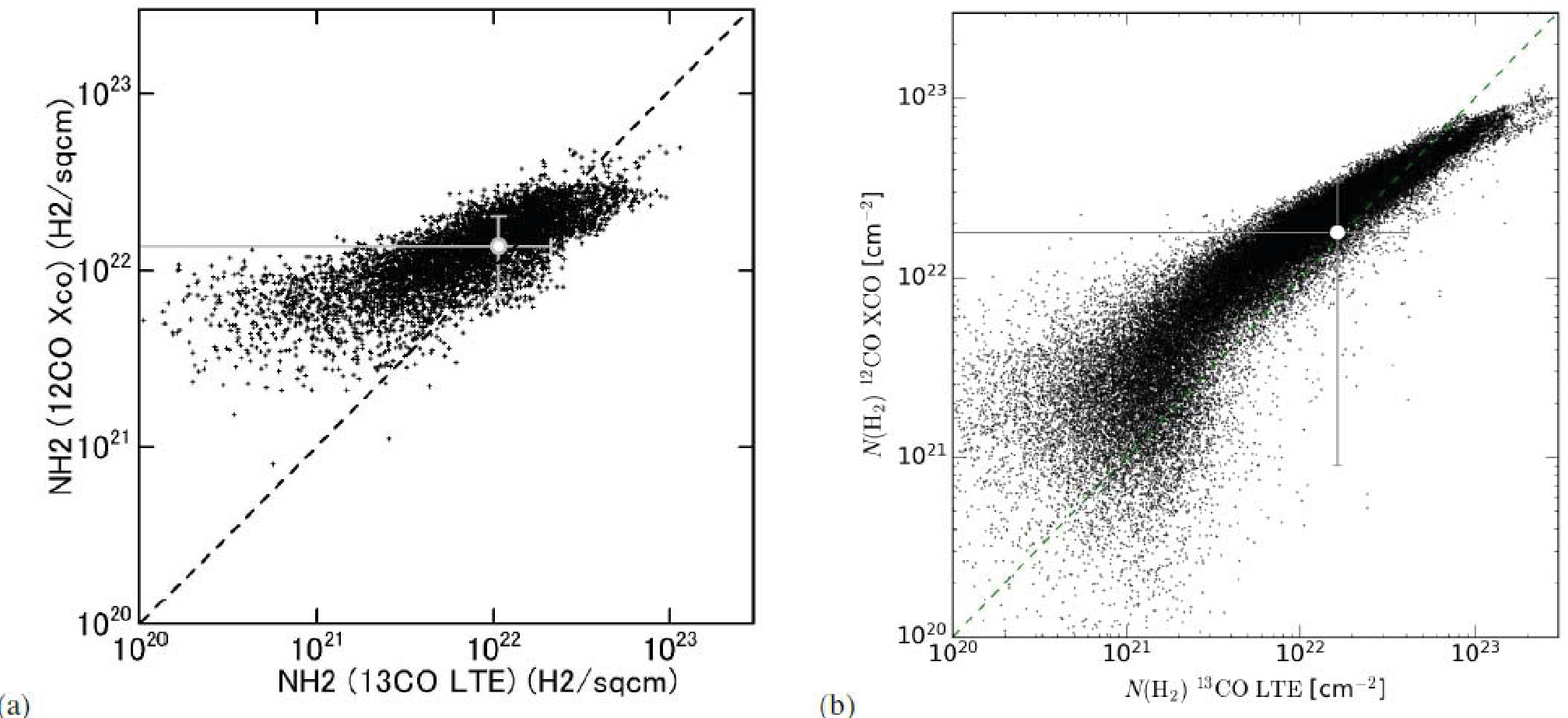}   
\end{center}
\caption{H$_2$ column density calculated using the $\Xco$ factor for the \co line intensity plotted against that calculated in the LTE assumption for \coth line intensity in the \revone{(a) M16 and (b) W43 region}. 
\revone{Here and hereafter, the dots represent values corresponding to all the individual grid points in figures \ref{maps} at $8''.5\times 8''.5$ interval. (For M16 dots are thinned out every 5 points.)}}
\label{NHtwo_m16w43} 
\end{figure*}

We then examine if the $\Xco$ and LTE methods yield identical results or not in different places in a single region or a cloud.
For this we calculate the column densities of H$_2$ in each cell (grid) of integrated intensity maps of the M16 and W43 regions. The excitation temperature and optical depth are calculated in each cell of the channel maps at a fixed velocity of the center channel in the used cube. Figure \ref{maps} shows the integrated intensity maps of the analyzed regions in \co line.

Figure \ref{NHtwo_m16w43} shows plots of the calculated \htwo column densities using the $\Xco$ and  LTE methods in individual cells of the M16 and W43 regions. 
The plots show a non-linear growth of curve in the sense that the $\Xco$ method yields a saturated values compared to LTE method. The same property has been reported in the Ophiucus and Perseus molecular clouds \citep{2008ApJ...679..481P}.

The big circles indicate averaged values of the plotted column densities with the bars denoting standard deviations of the plots. Despite of the large scatter in both axis directions, the averaged values from $\Xco$ and LTE methods agree with each other. We also superpose them on figure \ref{NHtwo_all}, where both the averaged points for M16 and W43 fall near the global linear line for the GMCs.

%%%%%%%%%%%%%%%%%%     

%%%%%%%%%%%%%%%%%%%%%%%%%%%%%%%%%%%%%%%%%%%%%%%%%%%%%
\section{Spectral Column Density (SCD)}
%%%%%%%%%%%%%%%%%%%%%%%%%%%%%%%%%%%%%%%%%%%%%%%%%%%%%
\subsection{Spectral (Differential) Column Density}
In this section, we examine more detailed relationships among various quantities such as the column densities derived by using the $\Xco$ and \coth-LTE methods. 
We first introduce a quantity to represent the column density corresponding to unit radial velocity (frequency), which we call the spectral, or differential, column density (SCD). 

The SCD for \co line brightness temperature is defined by
\begin{equation}
SCD_{\rm 12X} = \frac{d \NHtwo({\rm ^{12}CO})}{dv} 
=\Xco \Tbcotw 
\label{SCD12X}.
\end{equation}
The SCD for \coth line is defined by 
\begin{equation}
SCD_{\rm 13L}= \frac{d\NHtwo({\rm ^{13}CO})}{dv}  
= \Xcoth  \Tbcoth,
\label{SCD13L}
\end{equation}
where $\Xcoth$ is given by Eq. (\ref{eqXcoth}) including the function $Q$.
Here, $SCD$ is measured in \scdunit, 
$\Tb$ in [K], and $\Tex$ is calculated using Eq. \ref{Tex} from \co line brightness.

\begin{figure*}%%%%%%%%%%%%%%%%
\begin{center}           
{\bf \Large ----- M16 -----} \\      
\includegraphics[width=14cm]{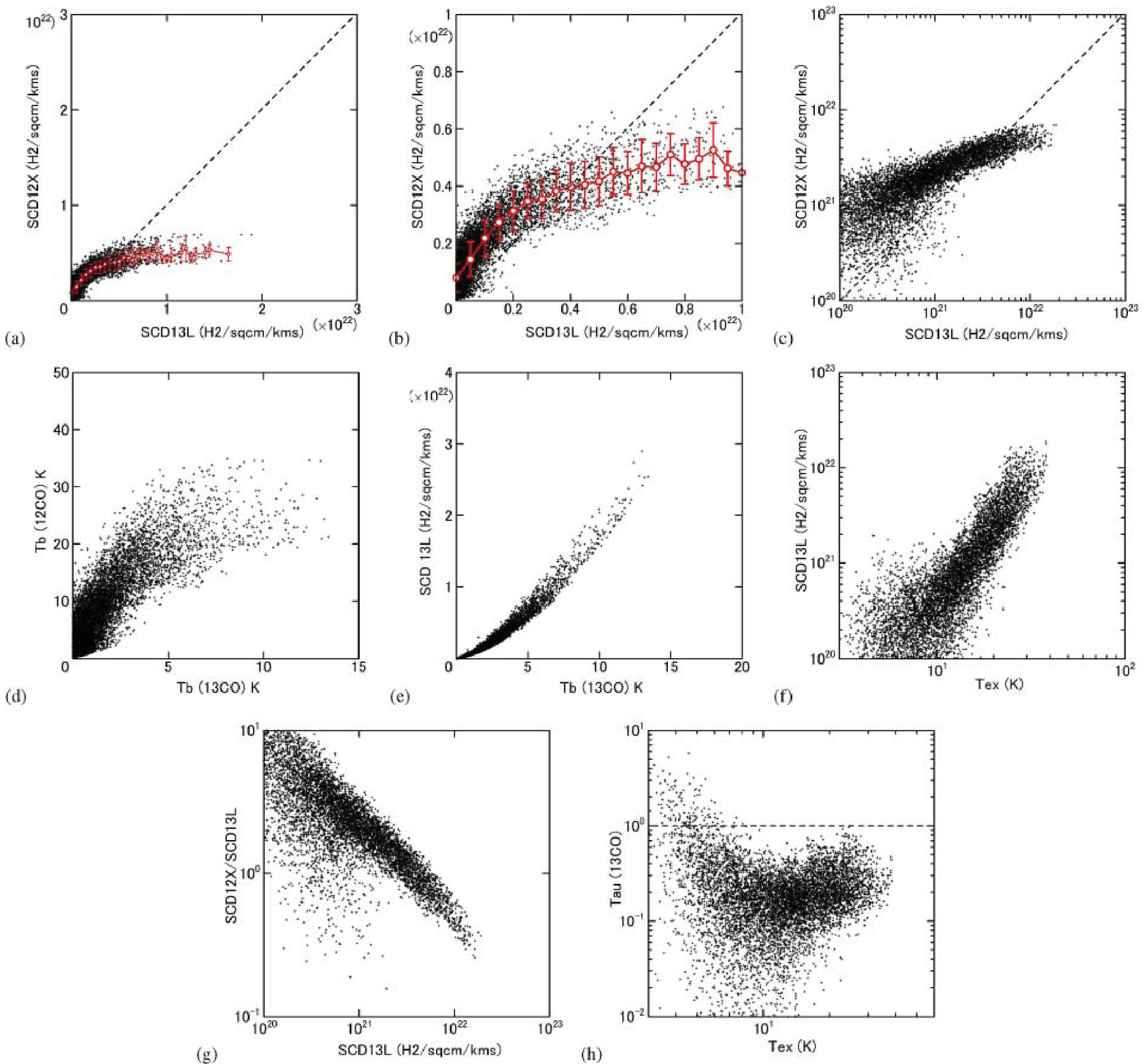}  
\end{center}
\caption{
Various plots in the $0\deg.5\times 0\deg.5$ region around the Pillars of Creation in M16 centered on G17.0+0.75 at $\vlsr=25$ \kms:
(a) Spectral column density, \scdx, against \scdl.  
\scdx underestimates the column at \scdl $> 4\times 10^{21}$ H$_2$ cm$^{-2}$ (\kms)$^{-1}$. 
(b) Same, but close up.
(c) Same, but in log-log plot.
(d) $\Tb$ of \co against \coth at $\vlsr=25$ \kms (TT plot).
(e) \scdl against $\Tb$($^{13}$CO). 
(f)  \scdl against $\Tex$.   
(g) Ratio \scdx/\scdl \scdl.
(h) Optical depth $\tau$($^{13}$CO) against $\Tex$, showing that the region is  almost optically thin in \coth.
\revone{Here and hereafter, red points denote Gaussian averaged values around each fixed abscissa points with full width equal to the interval, and the bars are standard deviations of vertical values in individual abscissa Gaussian bins. }
}
\label{fig_m16} 
\end{figure*}

\begin{figure*}%}%%%%%%%%%%%%%%%%
\begin{center}    
{\bf \Large ----- W43 -----} \\    
\includegraphics[width=14cm]{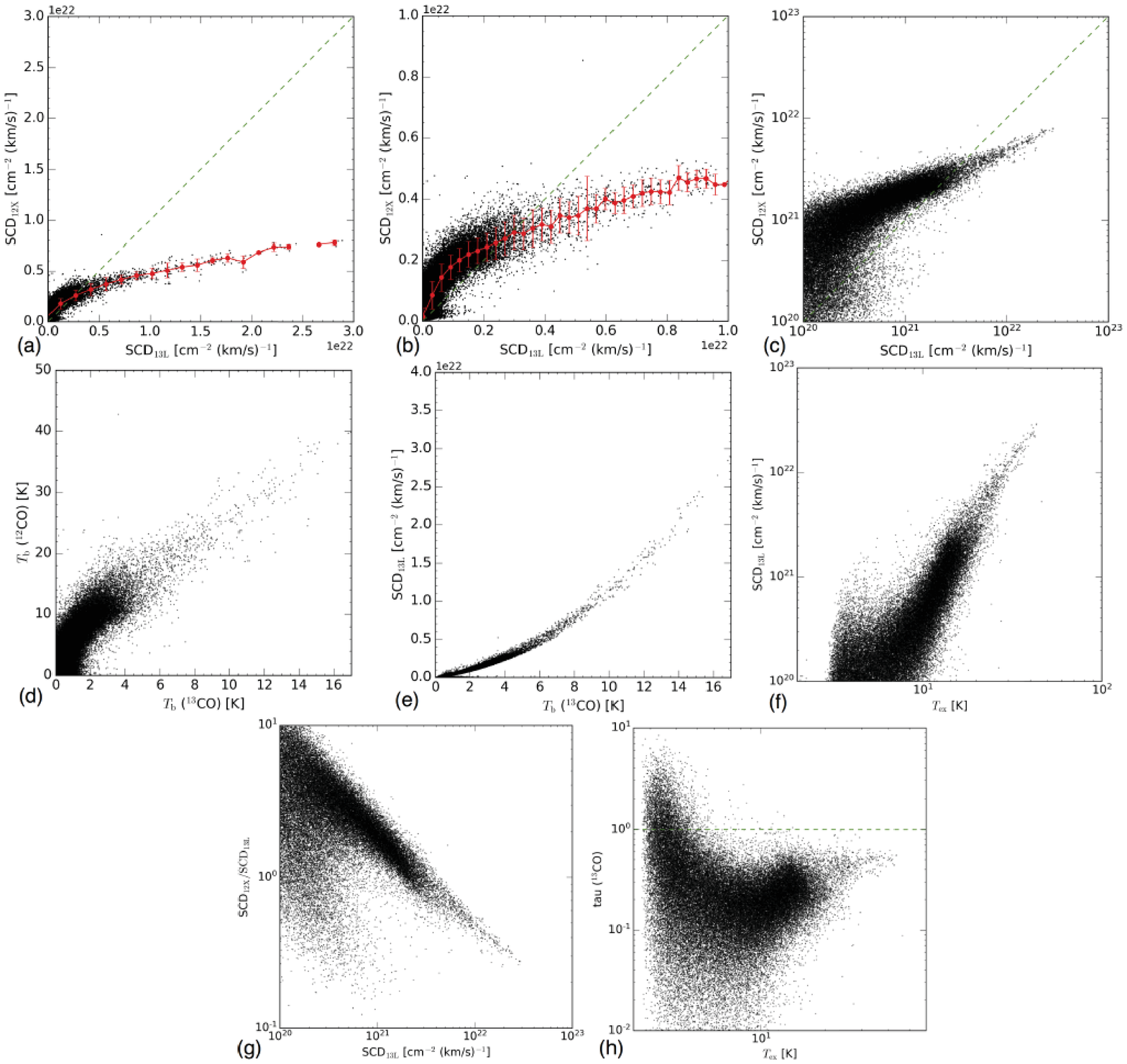}   
\end{center}
\caption{Same as figure \ref{fig_m16}, but for the $3\deg.0 \times 2\deg.0$ region around \revone{the W43 region} at $\vlsr=100$ \kms. 
(a) SCD, \scdx, against \scdl.
(b) Close up of SCD
(c) SCD in log-log plot. 
(d) TT plot. 
(e) \scdl against $\Tb$($^{13}$CO).
(f) \scdl against $\Tex$. 
(g) Ratio \scdx/\scdl against \scdl.
(h) Optical depth $\tau$($^{13}$CO) against $\Tex$.   }
\label{fig_w43} 
\end{figure*}

In figures \ref{fig_m16}(a-c)  we plot calculated \scdx against \scdl in individual cells in a channel map of M16 Pillar region at the line-center velocity, $\vlsr=\sim 25$ \kms, using the data from \cite{2020MNRAS.492.5966S}.  Figure \ref{fig_w43}(a-c) show the same for the \revone{W43 region} at $\sim 100 $ \kms using data from \cite{Kohno+2020}. 
The figures show significant saturation in \scdx, when \scdl exceeds a critical value at \scdl$\ge SCD_{\rm c}\sim 4\times 10^{21}$ \scdunit. 
 
 %%%%%%%%%%%%%%%%%%%
 
%}%%%%%%%%%%%%%%%%%%    

The majority of the points appearing in the bottom-left corner with large scatter indicate low-brightness and almost empty regions surrounding molecular clouds, whereas the high SCD and $\Tb$ cells in the top-right area represent those for dense clouds and cores.
The circles with bars show Gaussian running average of the plotted values in equal interval in the horizontal axis, and the bars denote the standard deviation in each interval.  

We may consider that \scdl using the optically-thin \coth line is a more natural tracer of the true column density. 
Then, the saturation in the vertical axis in the SCD plots  would indicate that the $\Xco$ conversion does not represent, or significantly under-estimate, the column density, when \scdl exceeds the critical value. 

\subsection{Various plots}

During the course, we also obtained various plots among the other quantities such as $\Tb$ and $\Tex$ in both lines.

Figures \ref{fig_m16}(d) and \ref{fig_w43}(d) show TT plots (scatter plots) between $\Tb$ of the \co and \coth lines in the same regions. The plot shows global correlation similar to the TT plot obtained for the entire Galaxy by \cite{2010PASJ...62.1277Y}, who reports decreasing slope with increasing temperature. This suggests that the curved property of the plot is a universal phenomenon.  

Figures \ref{fig_m16}(e,f) and \ref{fig_w43}(e,f) present the dependence of \scdl on the brightness and excitation temperature, indicating its increase with increasing temperatures. 
Figures \ref{fig_m16}(g) and \ref{fig_w43}(g) show that the ratio of \scdx to \scdl decreases with increasing column density.
Finally, Figures \ref{fig_m16}(h) and \ref{fig_w43}(h) plot the optical depth of \coth line against the excitation temperature, confirming that $\tau$ is sufficiently small in the regions with $\Tex$ higher than several K.

%%%%%%%%%%%%%%%%%%%%%%%%%%%%%%%%%%%%%%%%%%%%%%%%%%%%%%%%%%%%%%  
\section{\revone{Modified conversion relation} } 
%%%%%%%%%%%%%%%%%%%%%%%%%%%%%%%%%%%%%%%%%%%%%%%%%%%%%%%%%%%%%%

\revone{\subsection{New determination of conversion factor}}

\begin{figure*}%%%%%%%%%%%%%%%% 
\begin{center}    
\includegraphics[width=10cm]{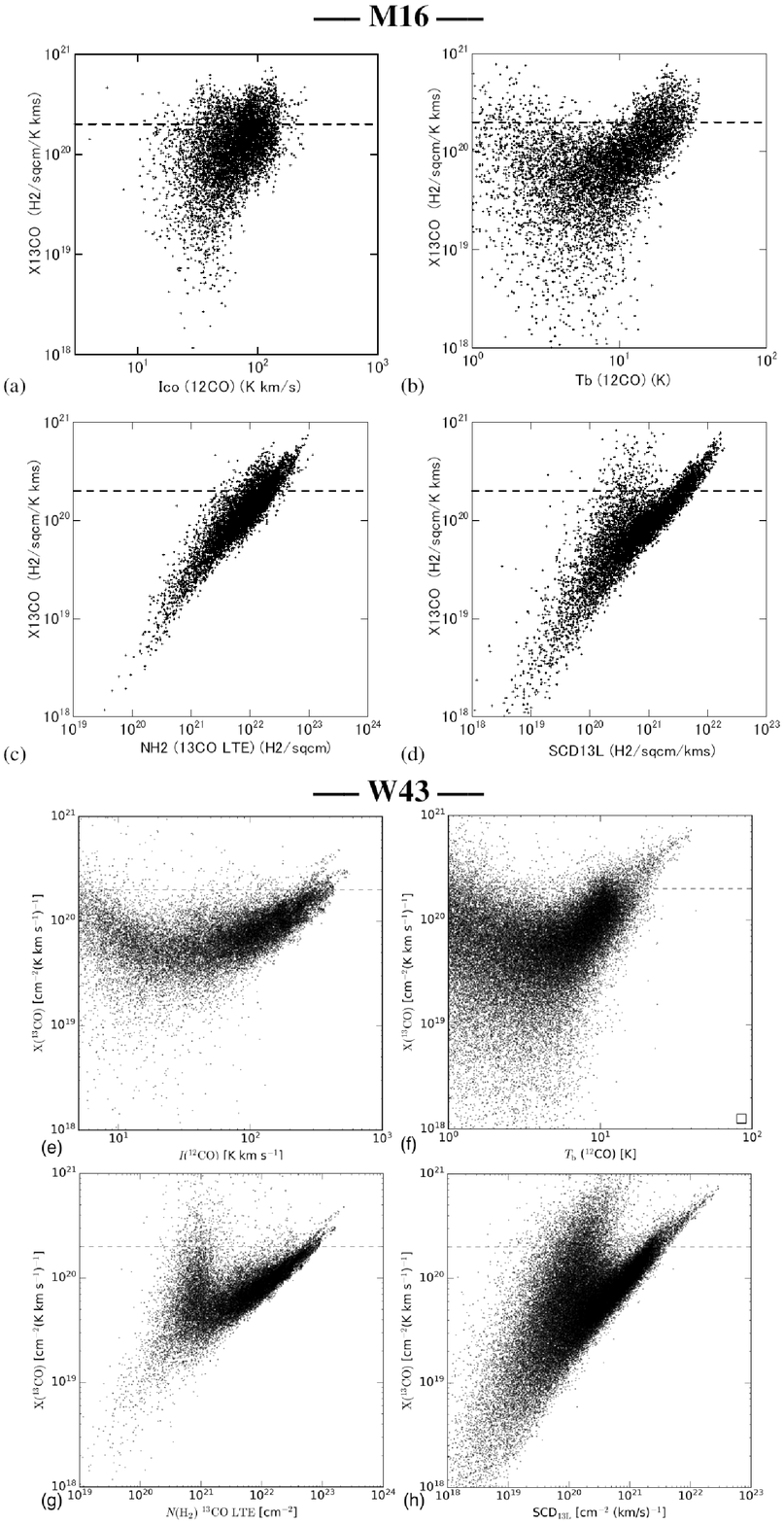}  
\end{center}
\caption{\revone{Factor $\Xcross (\equiv X_{\rm ^{13}CO}$: only in this figure) defined by Eq. (\ref{eqXnew}) plotted against $\Ico$ and $\Tb$ in \cotw, $N_{\rm H_2}$  and SCD  by \coth in LTE for the M16 (a-d) and W43 regions (e-h), showing that the "true" $X_{\rm CO}$ factor varies with the line intensity and column density.
Note that the V shaped features are due to the peculiar behavior of the function $Q$ of Eq. (\ref{Q}) }}
\label{newXco}
\end{figure*}

\revone{The newly introduced conversion factor, $\Xcross$, which relates the \Htwo column density calculated from \coth LTE measurement to the \cotw line intensity, defined by Eq. (\ref{eqXnew}) is useful to estimate a more reliable column density of \Htwo, even if there exist only \cotw measurements.
Figures \ref{newXco}(a,e) and (b,f) show plots of $\Xcross$ against $\Ico$ and $\Tb$ observed in the \cotw line emission, respectively, indicating that the conversion factor is an increasing function of the integrated \cotw intensity and brightness temperature.}

\revone{The plots are consistent with the similar plots obtained from the $A_v$ methods, even including the upward turnover at low intensities (column, $A_v$) 
\citep{Lee+2014}.
However, the here found general trend of increase in $\Xco$ with the column is quite contrary to the decreasing behavior as found by the $\gamma$-ray method in  anti-center outer Galactic clouds \citep{Remy+2017}.}

\revone{Figures \ref{newXco} (c,g) and (d,h) are the same plots against $\NHtwo$ and $SCD$ observed in the \coth line emission, respectively, which show that the conversion factor sensitively depends on the column density from \coth LTE and on the spectral column density (SCD). }

\revone{The plots show that the usage of a constant $\Xco$ overestimates the column density in low intensity or low density regions and clouds, whereas it underestimates at high column or intensity regions and clouds.
This result is consistent with, or rather equivalent to the result in the previous subsection. }

\revone{Although it may be possible to get a modified conversion factor as a function of $\Ico$ using figures \ref{newXco}(a), the scatter, particularly for M16, is too large to get conclusive fits.
Instead, in the following subsection, we will try to get a more reliable modification of the conversion law by fitting to similar plots in the spectral regime using SCDs.}

%%%%%%%%%%%%%%%% 

\subsection{Modified conversion relation}
\label{secmodify}

Being aware of the limitation and uncertainty of the $\Xgeneral$ business, we finally try to propose an approximate way to correct for the conversion factor.
We utilize the plots of the SCD for $\Xco$ and LTE methods in figures \ref{fig_m16}(a-c) and \ref{fig_w43}(a-c) to find the correction factor, assuming that \scdl represents the most reliable SCD. 

\begin{figure}%%%%%%%%%%%%%%%% 
\begin{center}        
\includegraphics[width=7cm]{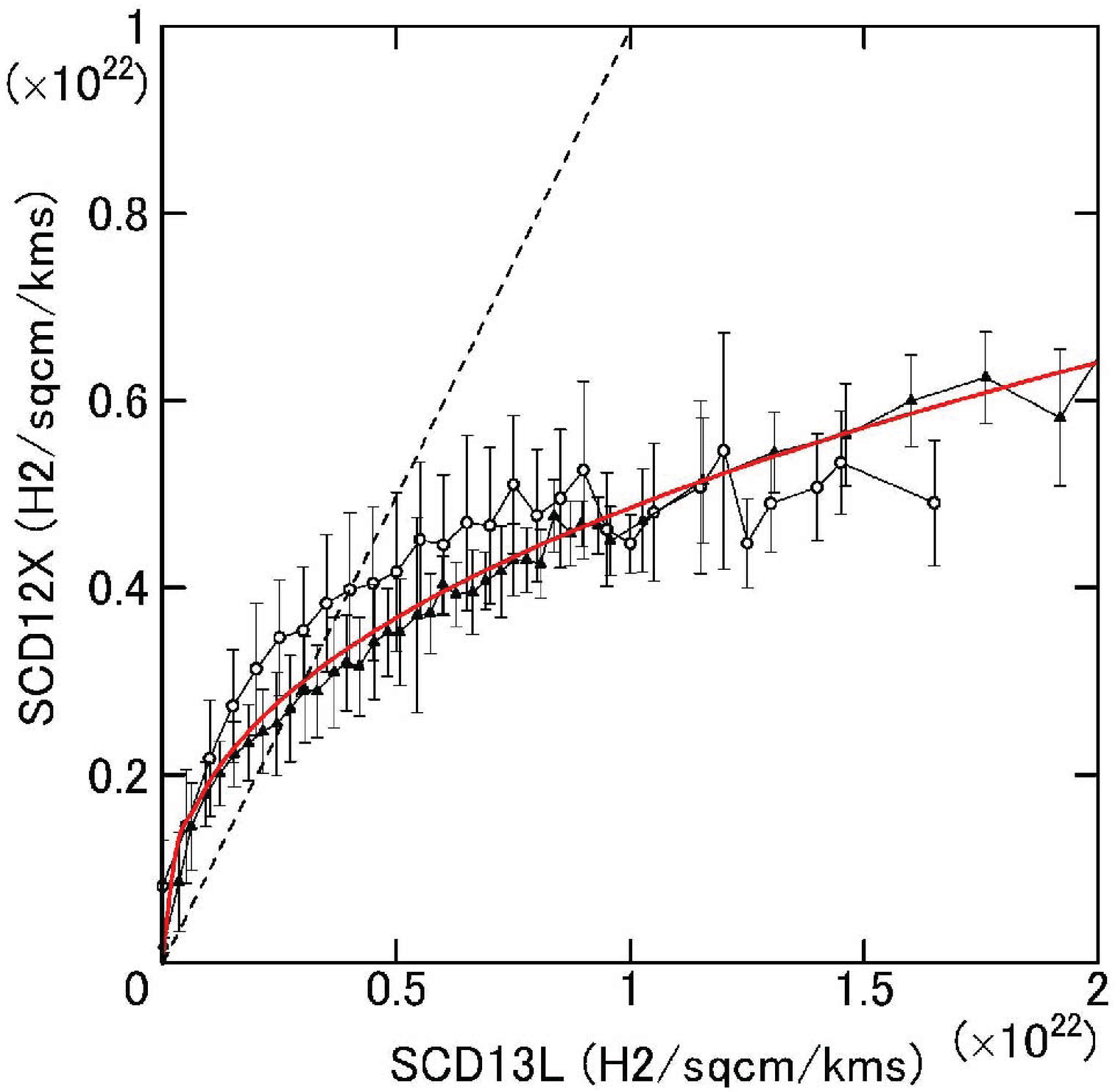}        
\end{center}
\caption{Running averaged \scdx vs \scdl plots for M16 (circles) and W43 (triangles), which are approximately fitted by a curve expressed by  
$\scx \sim SCD_{\rm c}\left( \frac{\scl}{SCD_{\rm c}} \right)^{\alpha} $ with $\alpha \sim 0.3-0.5$ and $SCD_{\rm c}\sim 2-4\times 10^{21}$ \scdunit.   } 
\label{m16w43dndv}
\end{figure}

In figure \ref{m16w43dndv} we reproduce the running averaged SCDs. The two curves for M16 and W43 coincide with each other within standard deviation.
We now try to fit the plots by a function as simple as possible, and propose a curve expressed by
\be
\scx  = SCD _{\rm c} \left(\frac{\scl}{SCD _{\rm c}}\right)^\alpha,
\ee
where $SCD_{\rm c} \sim 2-4\times 10^{21}$ \scdunit and $\alpha \sim 0.3-0.5$.
 The critical SCD is related to the critical brightness temperature of \co line as ${\Tb}^* ({\rm ^{12}CO})\sim SCD_{\rm c}/\Xco \sim 12-16$ K  in this figure. 
 
Rewriting $1/\alpha -1=\beta$, we obtain an approximate correction for the conversion factor as a function of the brightness temperature, $\Tb (=\Tb \ ({\rm ^{12}CO})$, hereafter), as
\be
\Xcomod(\Tb) = \Xco \left( \frac{\Tb}{ {\Tb}^*} \right) ^\beta,
\label{eqXmod}
\ee 
and a corrected formula to calculate the \Htwo column density using only the \co brightness temperature as follows:
\be 
SCD_{12X}^* \sim \Xco \left( \frac{\Tb}{\Tb^*} \right)^\beta \Tb ,
\ee
or
\be 
\NHtwo^* \sim \int \Xcomod(\Tb) \Tb dv
= \Xco \int \left( \frac{\Tb}{\Tb^*} \right)^{\beta} \Tb\ dv 
\label{eqNmod} 
\ee   
where  $\beta=1/\alpha -1 \sim 1-2$.

\revone{This is the fourth conversion formula obtained in this paper, which relates the spectral line profile of the \cotw line to a probable $\NHtwo$, as if it were determined by \coth-LTE measurement.
It corrects for the under-estimation by $\Xco$ method of $\NHtwo$ in dense clouds with higher brightness temperature than $\Tb^*$ or higher \scdl than $SCD_{\rm c}$, and does for the over-estimation in lower brightness or lower density regions. 
This formula is similar to the cross conversion relation given by Eq. (\ref{eqXnew}) against $\Ico$, but is more reliable in the sense that it is derived by the SCD analysis taking account of the variability of the (spectral) conversion factor as a function of $\Tb$.
, which varies with the radial velocity.}

%%%%%%%%%%%%%%%%%% 

\begin{figure*}%%%%%%%%%%%%%%%% 
\begin{center}    
\includegraphics[width=12cm]{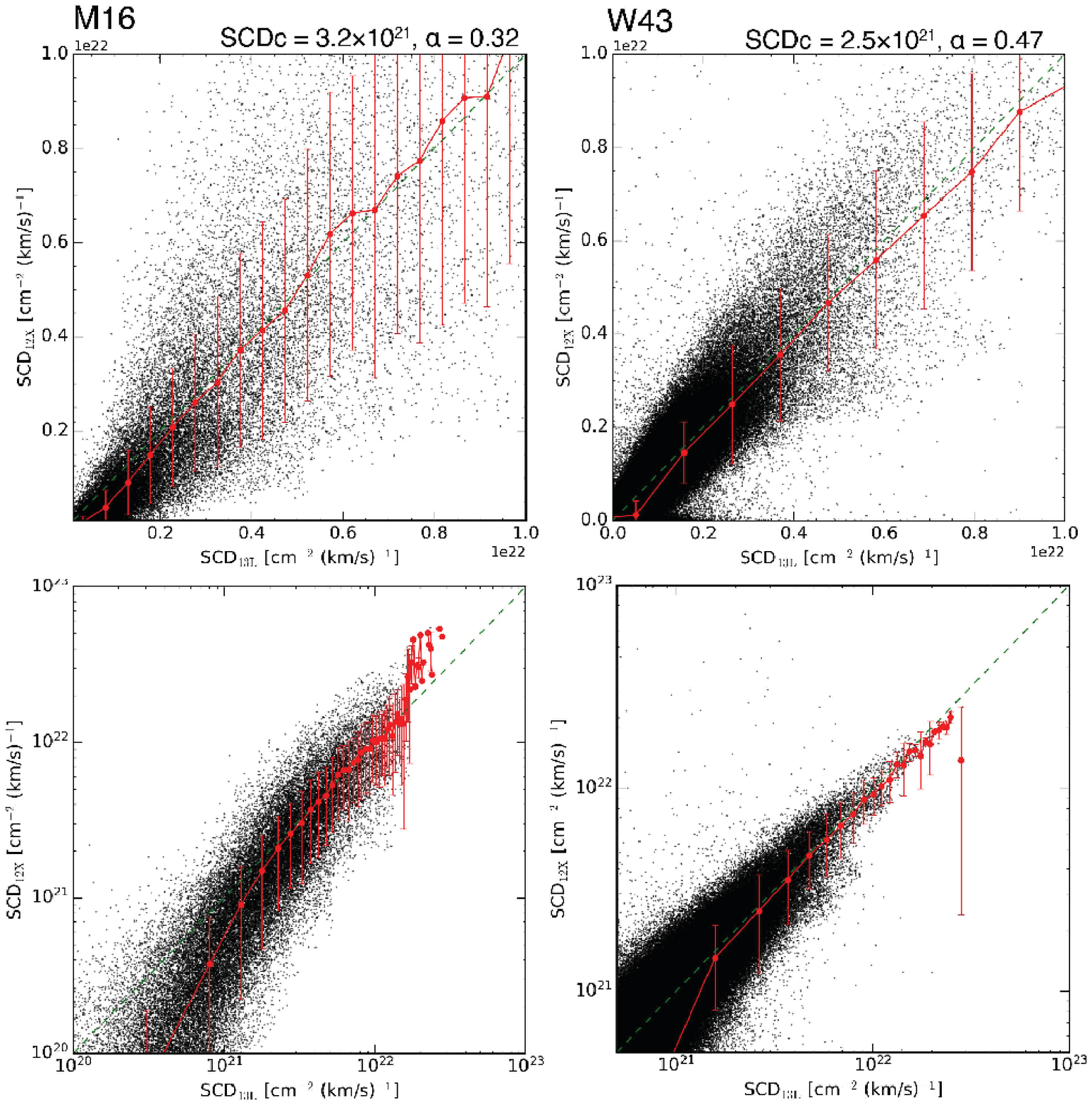}       
\end{center}
\caption{Corrected SCD relation for \revone{the M16 and W43 regions}. Note the improvement of the under- and over-estimation found in the original plots in figures \ref{fig_m16} and \ref{fig_w43} }
\label{c_scd}
\end{figure*}%%%%%%%%%%%%%%%%%% 

\begin{figure*}%%%%%%%%%%%%%%%% 
\begin{center}    
\includegraphics[width=12cm]{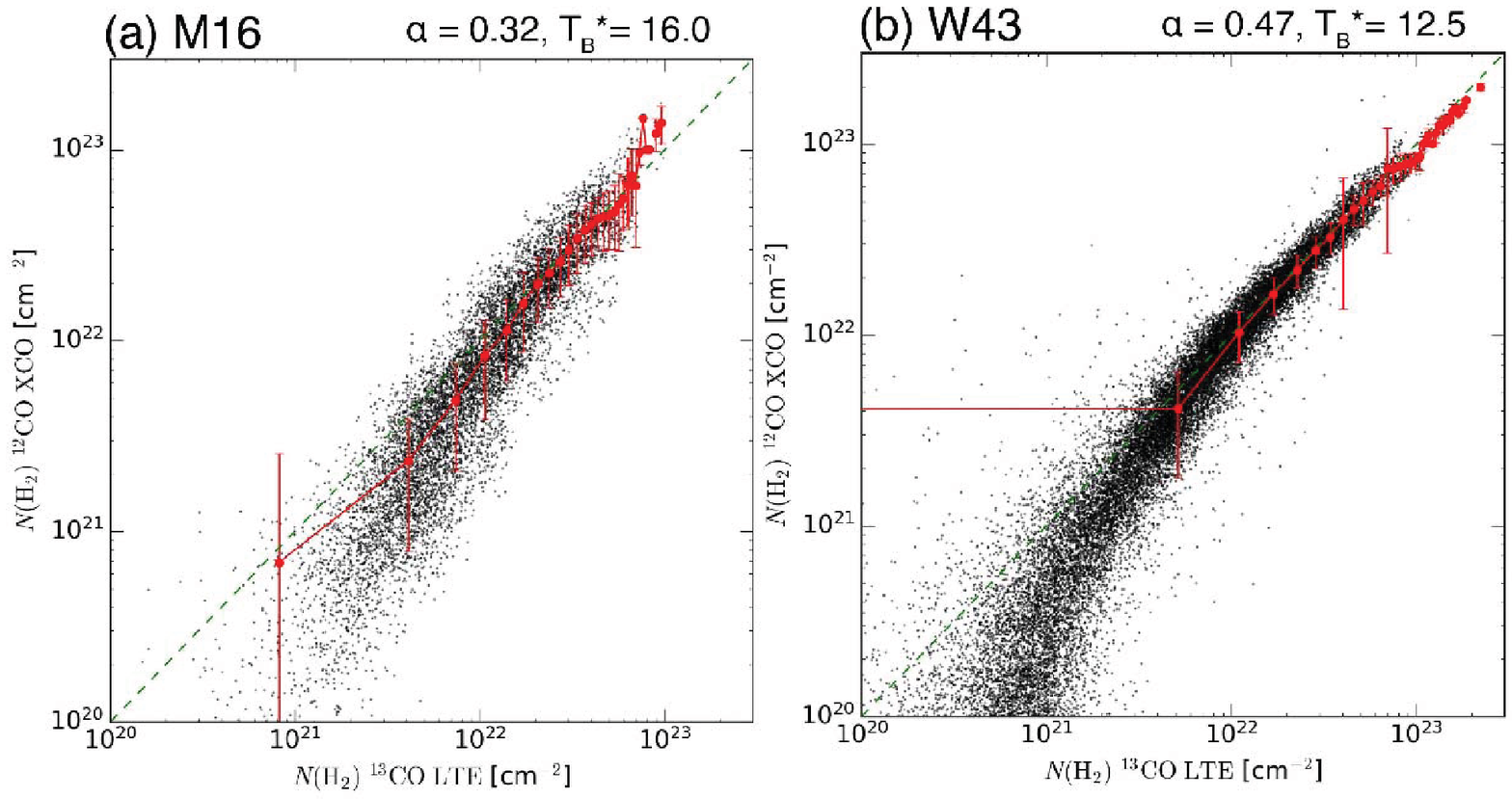}     
\end{center}
\caption{Corrected H$_2$ column density relation for \revone{the M16 and W43 regions}. Note the improvement of the under- and over-estimation found in the original plots in figures \ref{NHtwo_m16w43}(a) and (b). }
\label{cor_NH2}
\end{figure*}

In order to check if the correction works, we applied the modified conversion to the data of M16 and W43, and show the result for the SCD plots in figure \ref{c_scd}. As the best-fit parameters for M16, we obtained $\beta=2.1 \ (\alpha=0.32)$, $\Tb^* = 16.0$ K, and $SCD_{\rm c}=3.2\times 10^{21}$ \scdunit; and for W43 we obtained $\beta=1.13 \ (\alpha=0.47)$, $\Tb^*=12.5$ K, and $SCD_{\rm c}=2.5\times 10^{21}$ \scdunit. We also used $\Xcomod$ to calculate the H$_2$ column density for M16 and W43, and the result is shown in figures \ref{cor_NH2} (a,b). The displacements found in the original plots are largely reduced, so that the plotted points are distributed around the linear relations.

%%%%%%%%%%%%%%%%%% 

 It should be mentioned that the empirical relations are similar to each other between the MCs in M16 and W43 regions. This suggests that the relation is rather universal in MCs around SF regions, despite of the slightly different values of the critical values of $\Tb$ and SCD, which may depend on the properties of individual MCs such as the intensity of associated SF activity, distance from the GC, etc.. It would be a future subject to investigate the relation for a more different types of MCs such as those without SF regions, MCs in the GC, and/or those in the outer Galaxy.

%%%%%%%%%%%%%%%%%%%%%%%%%%%%%%%%%%%%%%%%%%%%%%%%%%%%%%%%%%%
\section{Discussion}
%%%%%%%%%%%%%%%%%%%%%%%%%%%%%%%%%%%%%%%%%%%%%%%%%%%%%%%%%%%
\subsection{Displacement of SCDs and self absorption}

We found that the spectral column density of H$_2$ molecules calculated using the $\Xco$ factor for the \co line is significantly displaced and underestimates the more realistic value of \scdl calculated for LTE assumption using the \coth line brightness.

This implies that the widely used $\Xco$ conversion may not hold in molecular clouds and local regions having higher SCD than the critical value with \scdl$\sim 4\times 10^{21}$ \scdunit. 

In addition to the saturation in \scdx at high column as the consequence of the condition that the \co line is optically thick, 
%it raises a problem about the assumption of the LVG (large velocity gradient) \citep{1983ApJS...51..203G},
%\citep{1974ApJ...187L..67S,1974ApJ...189..441G,1983ApJS...51..203G},
%which may not hold in high-column clouds.
the saturation is also accelerated by the factor $(1-\exp(-T_0/\Tex))^{-1}$ with $\Tex$, which approaches to $\Tex/T_0$ at sufficiently high $\Tx$, so that \scdl tends to $\propto \Tb \Tex\propto \tau \Tex^2$. Hence, both the saturation of the \co line and stronger dependence on $\Tex$ at high column will be the cause for the shallower increase of \scdx against \scdl plot.

\subsection{Integrated vs spectral, or global vs local conversion}
 
In figure \ref{NHtwo_all}, we plotted the column densities obtained from integrated intensities of the \co line using the $\Xco$ against those from the LTE assumption for \coth line in various GMCs.
The good linear correlation in the mean column and total mass would be due to the fact that the integration over the entire line profile smears out the fine line profiles, whose dip contributes only to a small fraction of the total intensity.  
The linear correlation would be also due to the averaging effect of spatially variable SCDs in each cloud. 

We may thus conclude that the $\Xco$ conversion using the \co line gives approximately the same column density as that calculated from the LTE method using \coth line, when it is applied to {\it integrated intensities averaged} over a cloud or region of scales from $\sim 10$ to $\sim 100$ pc. It is stressed that this statement evenly applies to such different regions as M16 and W43 with different properties and galacto-centric distances.

On the other hand, the $\Xco$ method significantly under-estimates the column density in higher density regions than the critical value,
\scdl$ \ge \sim 4\times 10^{21} $ \scdunit, 
or by column density, $\ge \sim 2\times 10^{22}$ [\Htwo cm$^{-2}$], but over-estimates in lower density regions. 

\subsection{Abundance ratio}

Throughout the paper except for section \ref{avNHtwo}, we adopted the \Htwo-to-$^{13}$CO abundance ratio of $\Ycoth=(5.0\pm 2.5)\times 10^5$ \citep{1978ApJS...37..407D}, and even smaller ratios are often employed \citep{2008ApJ...679..481P}. 
However, figures \ref{NHtwo_all} and \ref{NHtwo_m16w43}, where was adopted $\Ycoth=7.7\times 10^5$ \citep{Kohno+2020}, show a good agreement between the averaged values of $\NHtwo$ from $\Xco$ and LTE methods, as they lie on the dashed line showing that both are equal. On the other hand, if we adopt the lower abundance ($\Ycoth=(5.0\pm 2.5)\times 10^5$), the plots are shifted upwards by a factor of 1.5, significantly displacing from the dashed line. This might imply that a higher $\Ycoth$ is more plausible.
 
Since the LTE method is based on the line transfer of optically-thin \coth line, and hence simpler compared with the $\Xgeneral$ method based on various empirical plots of CO luminosity against the other \htwo mass tracers, 
we may consider that column density from LTE method would be more reliable, showing closer value to the true density. 

 \subsection{Limitation and uncertainty}
 
 We assumed that the \co line is optically thick, so that the excitation temperature is approximated by the brightness temperature. However, the self-absorption in \co line, which is rather common in dense clouds \citep{1981ApJ...245..512P}, would under-estimate $\Tex$.  This would affect the analysis including $\Tex$, particularly in the regions with high \scdl and $\Tex$, of the LTE method would still under-estimate the density. 

 On the other hand, in an opposite extreme case with low gas density, where the \co line is optically thin, the 'thick' assumption yields under-estimated $\Tex$. 
 This could be one of the reasons for the over-estimated column density by the $\Xco$ method at low column regions. 

\begin{figure}%%%%%%%%%%%%%%%% 
\begin{center}        
\includegraphics[width=8cm]{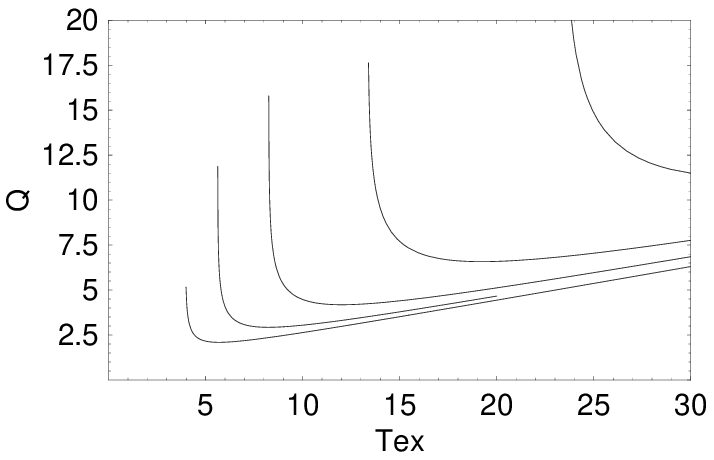}   
\end{center}
\caption{$Q$ function plotted against $\Tex$ (K) for $\Tbcoth=1, 2.5$, 5, 10, and 20 K (from bottom to top curve). The function accounts for the V shaped behavior in figure \ref{newXco}. } 
\label{Qfunc}
\end{figure}

As shown in figure \ref{Qfunc}, function $Q\left(\Tex,\Tb(^{13}{\rm CO})\right)$ (equation \ref{Q}) tells us that under-estimated $\Tex$ affects \scdl in a complicated way in such a way that the column is under-estimated at high $\Tex$ and over-estimated at low $\Tex$. This would cause additional scatter in the \scdl-\scdx\ plot.
The peculiar effect of the function is more directly observed in the V shaped behavior of the plots in figure \ref{newXco}.
 For more precise measurement of gas density in such core regions, a more sophisticated analysis including the line transfer would be necessary, while it is beyond the scope of this paper.

%%%%%%%%%%%%%%%%%%        

\revtwo{As we used the \cotw and \coth line data, the result cannot be applied to  clouds in CO possibly present in the form of either extremely low temperature molecules, dust, HI, higher-temperature gases than $\sim 100$K including plasma, or CO-dark regions such as PDR (photo-dissociation regions) \citep{Hollenbach+1997}.
On the other hand, the modified conversion relation, Eq. (\ref{eqNmod}), may be applicable to "CO-faint" clouds or regions, in so far as they can be detected by CO. }

\revtwo{The various relations obtained in the present analysis, employing \coth line data yield uncertainty of the same order as that of the abundance ratio of the $^{13}$CO molecule to the \Htwo gas, which is about a factor of $\sim 1.5$. }

\revtwo{Despite the limitations and uncertainties, the advantage of the method would be its applicability to small clouds and structures of sub-pc scales and low CO line brightness regions at $\Tb \ge \sim 1$ K, according to the high angular resolution ($20''$) of the observed data and noise temperature ($\sim 1-1.5$ and $0.7-0.9$ K in \cotw and \coth, respectively) for M16 at a distance of 2 kpc \citep{Guarcello+2007} and W43 at 5.5 kpc \citep{Zhang+2014}.}

\revone{\subsection{The variability of conversion factor}}

\revone{We have shown that, when it is applied in individual directions within a molecular cloud, the widely accepted conversion factor, $\Xco\sim 2\times 10^{20}$ \xcounit, combined with the $^{12}$CO intensity significantly under-estimates the column density in the $^{12}$CO-opaque cloud cores and high-intensity regions.
On the contrary, it over-estimates the column in the envelopes and inter-cloud regions having low density and weak line intensities.
Namely, since extended objects such as cloud complexes and associations are spatially dominated by low-brightness regions, their total masses tend to be over-estimated for the increasing-area effect, when they are integrated over the entire area.
Such a problem of over- or under-estimation may be solved by applying the modified conversion relations like Eq. (\ref{eqXnew}), or equivalently using Eq. (\ref{eqNmod}).
We summarize the various conversion relations discussed in this paper along with  their merits and demerits in their usage in table \ref{tabXco}.\\}

\begin{table*}
\begin{center}
\caption{\revone{Various conversion factors and relations.} } 
\label{tabXco}
\begin{tabular}{llll} 
\hline\hline\\
Method & Eq. & Formula$^\dagger$   & Remarks\\
\hline
(1) Direct $^{12}$CO 
&(\ref{NHtwoXco})
& $\NHtwo=\Xco \Icotw =2\times 10^{20}\Icotw$
& Simple; Sensitive; No need $^{13}$CO;\\ 

&
&
& Over/under at low/high columns.\\ 

&&&\\

(2) Direct $^{13}$CO, LTE
&(\ref{NHtwoLTE})
&$\NthLTE=\Xcoth  I_{\rm ^{13}CO} =1.50 \times 10^{20} Q \Icoth$
&Accurate; Need deep $^{13}$CO map.\\  

&&&\\ 

(3) Cross; Intensity
&(\ref{eqXnew})
&$\NtwthLTE=\Xcross \Icotw$  
&Moderate; Hard to fit Fig. \ref{newXco}(a).\\ 

&&&\\

(4) Modified; Spectral 
&(\ref{eqNmod})
&$\NHtwo^{*}=\Xco \int \left(\frac{\Tb}{\Tb^*}\right)^{\beta}\Tb\ dv$  
&Accurate; Sensitive; Need $^{12}$CO $\Tb(v)$ cube.\\

\hline
\end{tabular}  \\  
$^\dagger$ Numerics in unit of \xcounit.
\end{center}
\end{table*}

%%%%%%%%%%%%%%%%%%%%%%%%%%%%%%%%%%%%%%%%%%%%%%%%%%%%%%
\section{Summary}
%%%%%%%%%%%%%%%%%%%%%%%%%%%%%%%%%%%%%%%%%%%%%%%%%%%%%%

CO-to-\Htwo conversion using the constant conversion factor, $\Xco\sim 2\times 10^{20}$ \xcounit, gives reasonable estimation of the \Htwo column density in molecular clouds, only when it is applied to estimation of the averaged integrated intensity over the cloud and of total molecular mass.
However, the $\Xco$ method significantly underestimates the molecular density in dense clouds and local regions having \scdl greater than the critical value of $\sim 3\times 10^{21}$ \scdunit, which is understood as due to self absorption of the \co line in dense and high $\Tex$ regions. On the contrary, it over-estimates in lower density regions than the critical value.
\revone{This implies that the specific conversion factor is dependent on the gas density and line intensity, rising in the opaque cloud cores with high line intensities, and decreasing in the envelopes and inter-cloud regions. }

\revone{Assuming that the LTE method using \coth line gives more reliable estimation of the \Htwo column density, and based on the empirical fitting to the \scdx-\scdl plot, we proposed a modified (spectral) conversion factor given by Eq. (\ref{eqXmod}), and a new conversion relation given by Eq. (\ref{eqNmod}).
The new formula corrects for the over/under estimation in cloud envelopes/cores, and yields reliable $\NHtwo$, even if we have only \cotw line data.}

\section*{Acknowledgements}

Data availability: This paper made use of the data taken from the FUGIN project 
(\href{http://nro-fugin.github.io}{http://nro-fugin.github.io}).
The FUGIN CO data were retrieved from the JVO portal (\href{http://jvo.nao.ac.jp/portal}{http://jvo.nao.ac.jp/portal}) operated by ADC/NAOJ.
The Nobeyama 45-m radio telescope is operated by the Nobeyama Radio Observatory, and the data analysis was carried out at the Astronomy Data Center (ADC) of National Astronomical Observatory of Japan.
We utilized the Python software package for astronomy \citep{2013A&A...558A..33A}. 
The authors are grateful to the anonymous referee for the useful comments.

\end{document}